\documentclass[%
 reprint,
 amsmath,amssymb,
 aps,
prb,
longbibliography,
]{revtex4-1}
\setcitestyle{numbers,square}

\usepackage{physics}
\usepackage{graphicx}
\usepackage{dcolumn}
\usepackage{bm}
\usepackage[normalem]{ulem} 
\usepackage{color}

\usepackage{hyperref}
\hypersetup{
    colorlinks,%
    citecolor=blue,%
    linkcolor=blue,%
    urlcolor=blue
}

\newcommand{\fp}{\it first-principles}
\newcommand{\avg}[1]{\langle #1 \rangle}

\newcommand{\mean}[1]{\expval{#1}} 

\begin{document}

\title{Orbital pseudospin-momentum locking in two-dimensional chiral borophene}

\author{F. Crasto de Lima} 
\author{G. J. Ferreira}
\author{R. H. Miwa}

\affiliation{Instituto de F\'isica, Universidade Federal de Uberl\^andia, \\ C.P. 593, 38400-902, Uberl\^andia, MG,  Brazil}

\date{\today}

\begin{abstract}
Recently, orbital-textures have been found in Rashba and topological insulator (TI) surface states as a result of the {spin-orbit coupling (SOC)}. Here, we predict a $p_x/p_y$ orbital texture, in linear dispersive Dirac bands, arising at the K/K' points of $\chi$-$h_0$ borophene chiral monolayer. Combining {\fp} calculations with effective hamiltonians, we show that the orbital pseudospin has its direction locked with the momentum in a similar way as TIs' spin-textures. Additionally, considering a layer pseudospin degree of freedom, this lattice allows stackings of layers with equivalent or opposite chiralities. In turn, we show a control of the orbital textures and layer localization through the designed stacking and external electric field. For instance, for the opposite chirality stacking, the electric field allows for an on/off switch of the orbital-textured Dirac cone.
\end{abstract}

\maketitle

\section{Introduction}

Since the synthesis of graphene, two dimensional (2D) materials and related 2D states have become a major field of study in physics and materials science \cite{SCIENCENovoselov2004, NANOSCALEMas2011, SCIENCENovoselov2016}. Many interesting characteristics arise in this lower dimension, from intrinsic Dirac cone in graphene \cite{RMPCastro2009}, to the confinement induced 2D topological states in topological insulators (TI) \cite{PRLFu2007, RMPHasan2010}. Additionally, application of the spin texture of 2D surface states, arising due Rashba SOC or TI system, has been found prominent in the magnetic moment switching of ferromagnets, through the Edelstein effect and spin torque transfer \cite{SSCEdelstein1990, PRBSilsbee2001, PRBRodriguez2017, SCIREPGeng2017}. Further detailed analysis has shown that, within the TI spin-texture, an associated orbital texture exist, due to the strong SOC \cite{PRLZhang2013, NATUREWaugh2016, PRBGotlieb2017}. Such orbital texture was already experimentally observed\cite{NATUREXie2014,PRBMiao2014}.

Notwithstanding the spin texture, the existence of orbital texture allows for interesting phenomena. For instance (i) the orbital hall effect \cite{PRLGiant2009, PRBJo2018, PRLGo2018}, where an electric field induces a flow of orbital angular momentum in a transverse direction; (ii) the emergence of chiral orbital angular momentum due to an orbital Rashba effect \cite{PRLPark2011, PRBPark2012}; and (iii) electronics based in the orbital degree of freedom \cite{PRLBernevig2005, SCIREPGo2017, JPCMMing2018}. {While the spin Edelstein effect has been reported as a prominent phenomena for the design of magnetic memory,\cite{NATUREManipatruni2019} the introduction of an orbital degree of freedom has shown to increase the efficiency of torque transfer in such devices.\cite{NATUREChen2018}} Such pure orbital texture was shown to emerge in parabolic bands in the two-dimensional material BiAg$_2$ \cite{SCIREPGo2017}. Despite that, {to the best of our knowledge}, the emergence of a pure orbital-texture in linear Dirac bands, without the SOC, is still unexploited. Nonetheless, uncovering different orbital textures and ways to control it, allow for new possibilities in materials design.

Recently, the synthesis of borophene \cite{ScienceMannix2015, NatureChemFeng2016, PRLFeng2017}, a boron analogue of graphene, have increased the realm of 2D materials. Especially because boron atoms present different planar coordinations, which leads to the existence of many stable phases \cite{NanoRXu2016, JPCLYi2017, FPFong2018}. Interestingly, boron has shown to be stable in a Archimedean lattice that has two chiral variants, namely $\chi$-$h_0$ phases \cite{JPCLYi2017, MATBranko1977}. Timely studies have brought into attention that the chiral characteristic of the lattices leads to new topological phases in 3D materials \cite{PRLChang2017,NATUREChang2018}. Therefore, the exploration of its 2D counterpart can uncover novel phenomena.

In this paper, we discuss the existence of orbital pseudospin-texture in borophene $\chi$-$h_0$ lattices. This orbital texture, in contrast with the originated in TIs \cite{NATUREWaugh2016} and Rashba surface states \cite{PRBHanakata2017}, arises without spin-orbit coupling \cite{PSAlexandradinata2015}. Nevertheless, we find the presence of linear dispersive Dirac cone with orbital pseudospin-momentum locking, analogous to spinful TI surface states. Furthermore the existence of two chiral variants of this borophene phase allows for a design of the stacking sequence. Combining the orbital pseudospin with this layer degree of freedom, we show that its orbital-textures and layer localization can be controlled through an external electric field. Particularly, the orbital texture of the same chirality stacking is robust against the electric field. On the other hand, for stackings of layers with opposite chirality, the electric field allows for an on/off switch of the orbital-pseudospin Dirac cone.

\section{Results and Discussions}

\subsection{Monolayer orbital-texture}

\begin{figure}
\includegraphics[width=\columnwidth]{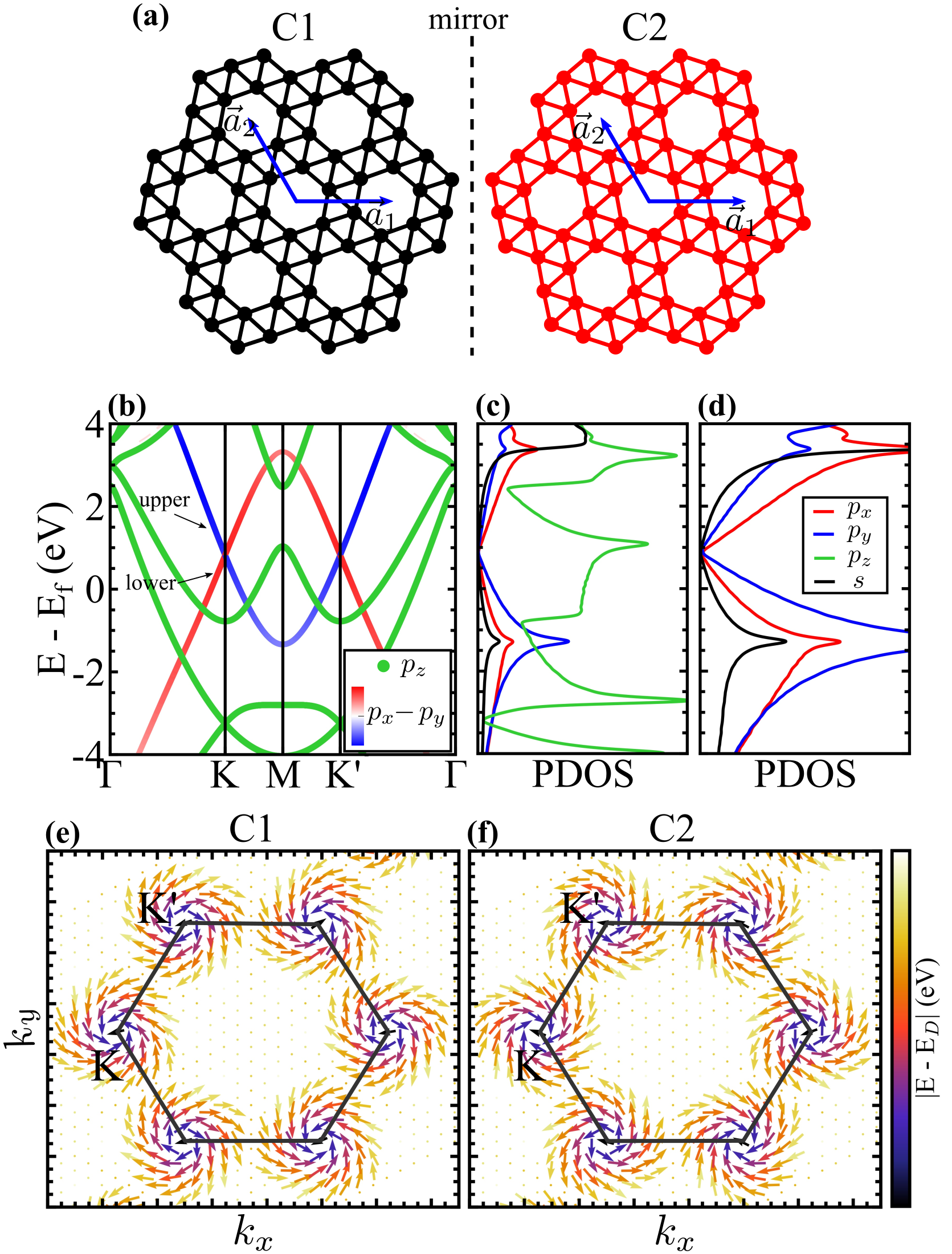}
\caption{\label{str-band} (a) Atomic structure of the borophene $\chi$-$h_0$ with its two possible chiralities C1 and C2. (b) Orbital projected band structure. (c) and (d) projected density of states. Orbital texture for the Dirac cone (e) borophene-C1 and (f) borophene-C2. The vectors are the mean value of the orbital pseudospin operators $(\mean{\hat{\sigma}_x},\,\mean{\hat{\sigma}_y})$, and the color code indicates the energy distance from the Dirac point.}
\end{figure}

The borophene $\chi$-$h$ phases were recently proposed and shown to have greater stability compared to its synthesized $\alpha$ phase \cite{JPCLYi2017}. In particular, among the $\chi$ phases, the $\chi$-$h_0$ was shown to be the most stable in the freestanding form. In this structure, Fig.\,\ref{str-band}(a), all boron atoms are five-fold coordinated, and arranged in a hexagonal lattice. Such borophene lattice arises in two chiral variants, mirror images of each other, which we will refer as borophene-C1 and borophene-C2 structures, Fig.\,\ref{str-band}(a). Both forms belong to the $P/6m$ ($C_{6h}$) space group. Based on the Density Functional Theory (DFT), we find a lattice parameter of $4.48$\,{\AA}, with mean first-neighbors distance of $1.69$\,{\AA}.

From the electronic perspective, the system is metallic with dispersive bands crossing the Fermi level, arising mostly from the $p_z$ and $p_x/p_y$ orbitals [Fig.\,\ref{str-band}(b)]. The $p_x/p_y$ bands form a Dirac cone at E$_D=0.82$\,eV above the Fermi energy at the $K$ and $K'$ points. Furthermore, the atomic orbital projected density of states, Fig,\,\ref{str-band}(c), presents the characteristic V-shaped behavior of a Dirac cone. A close inspection, Fig.\,\ref{str-band}(d), shows that the $s$-orbital contribution goes faster to zero near the Dirac point, and can be neglected. The linear dispersive band presents high electron velocity, $v_D=1.12\times10^{6}$\,{m/s}, of the same order of magnitude as graphene's \cite{RMPCastro2009}. Additionally, the $p_x/p_y$ cone presents a uneven distribution of the orbitals in the Brillouin zone (BZ), which will be further discussed below.

Due to the planar structure of this borophene phase, the states with different in-plane mirror eigenvalues are decoupled. For instance the $p$ orbitals can be separated into odd $p_z$ and even $p_x/p_y$ subspaces. Therefore, the $p_x/p_y$ orbitals form a pseudospin degree of freedom (composed by the $E'$ irreducible representation of the $C_{3h}$ group at $K$ or $K'$). Thus, we define spinor-like operators, $\sigma_x$, $\sigma_y$ and $\sigma_z$ constructed in the basis of $l=1$ and $m=\pm1$ orbitals, i.e. $\ket{\pm} = \left( \ket{p_x} \pm i\,\ket{p_y} \right)/\sqrt{2}$. Within the DFT calculations we compute the mean values of $\avg{\sigma_x}$ and $\avg{\sigma_y}$ pseudospinor operators and represent them in a vector field plot along the BZ for the upper band of the cone, Fig.\,\ref{str-band}(e) and (f). Near each Dirac cone ($K$ and $K'$ points) we see a resemblance with the spinful texture of 3D topological insulators. Notice that the textures at the $K$ and $K'$ points rotate in opposite directions within each lattice (C1 or C2), with C1's $K$-texture matching C2's $K'$-texture, \textit{c.f.} Fig.\,\ref{str-band}(e) and (f). Analogous behavior is seem in the lower cone branch, but with textures reversed in relation to the upper branch.

We can further explore this orbital pseudospin in a effective model. For the borophene-C1 at the $K$ point the helical orbital pseudospin hamiltonian can be written as
\begin{equation}
h^{(1)}_{K} = +h = \hbar v_D \left( q_x \sigma_y - q_y \sigma_x \right), \label{eq:h1}
\end{equation}
where $v_D$ is the Dirac velocity, $(q_x,\,q_y)$ are the momentum vector components measured from the $K$ point, and $(\sigma_x, \sigma_y, \sigma_z)$ are the Pauli matrices in the orbital pseudospin degree of freedom. Such hamiltonian has eigenvalues given by
\begin{equation}
e_{\lambda} = \lambda v_D |\vec{q}|, \label{eq:eig-h1}
\end{equation} 
with $\lambda=\pm 1$ for each of the two linear dispersive bands, and eigenvectors
\begin{equation}
\ket{e_\lambda} = \frac{1}{\sqrt{2}} \left( \ket{+} + \frac{\lambda\left( -q_y +iq_x \right)}{|\vec{q}|} \ket{-} \right). \label{eq:eiv-h1}
\end{equation}
{We can relate the $\lambda$ quantum number with the helicity of each states, as the proposed hamiltonian comutes with the orbital counterpart helicity operator\cite{PRLFerreira2013}.} From these eigenvectors, we calculate the orbital texture at each {$\vec{q}$}, yielding
\begin{equation}
\Big(\avg{\sigma_x}_{\lambda},\, \avg{\sigma_y}_{\lambda}  \Big) = \frac{\lambda}{|\vec{q}|} \left( -q_y,\,q_x  \right),
\end{equation}
which gives the {right-handed ($\lambda=1$) helicity} seeing in the DFT results of Fig.\,\ref{str-band}(e) at the $K$-point, with the orbital pseudo-spin orthogonal to the $\vec{q}$ vector. Furthermore, to obtain the Hamiltonian and orbital texture at the $K'$ point, we consider the inversion symmetry. The inversion takes $\vec{K} \rightarrow -\vec{K} \equiv \vec{K}'$ and $\vec{k} \rightarrow -\vec{k}$. Since $\vec{q} = \vec{K}-\vec{k}$, it also takes $\vec{q} \rightarrow -\vec{q}$. Moreover, since our basis is composed by $p_x$ and $p_y$ orbitals, the inversion operator in the Hilbert space becomes $I = -\sigma_0$. Therefore, the $K'$ Hamiltonian becomes $h^{(1)}_{K'}(\vec{q}) = I h^{(1)}_{K}(\vec{q}) I^{-1} = h^{(1)}_{K}(-\vec{q}) = -h$. This sign change leads to the counter rotating texture at the $K'$ point in relation to $K$ point, Fig.\,\ref{str-band}(e).

The borophene-C2 monolayer is obtained as the mirror of the C1 monolayer, see Fig.\,\ref{str-band}(a). On the reciprocal space, this mirror operation takes $K \leftrightarrow K'$. Accordingly, C2 can be described as a C1 monolayer with the $K$ and $K'$ points exchanged. Therefore the hamiltonian for  C2 at the $K$ point is, $h^{(2)}_{K} = -h$. Consequently, it has the same eigenvalues as $h^{(1)}_K$, but with eigenvectors leading to a reversed pseudospin texture, capturing the DFT results, \textit{c.f.} Fig.\,\ref{str-band}(e) and (f).

The momentum locked pseudospin texture leads to the orbital counterpart of Edelstein and inverse Edelstein effect \cite{SSCEdelstein1990, PRLShen2014}. For instance{, within the inverse Edelstein effect, the proportionality constant ($\lambda_{IEE}$) between a linear charge current density and a perpendicular areal orbital current density is defined by the Fermi velocity\cite{PRLRojas2016} ($\lambda^D_{IEE}=v_D \tau$) for materials with linear Dirac spectrum, and by the Rashba coefficient\cite{PRLRojas2016} ($\lambda^R_{IEE} = \alpha_R \tau /\hbar$) for the orbital Rashba effect induced orbital-texture.} Due to the high Fermi velocity found here ($v_D \sim 10^{6}$\,m/s), {we} expect that the inverse Edelstein induced current can be {$\lambda^{D}_{IEE}/\lambda^{R}_{IEE} \sim 6.5$} times larger than the one in the orbital Rashba bands of BiAg$_2$\cite{SCIREPGo2017}, considering similar momentum relaxation time ($\tau$). {Note, however, that in borophene the $p_z$ metallic bands will have a contribution in experimental transport measurements. Moreover, despite not having a trivial tuning capability, the orbital pseudospin degree of freedom can be experimentally observed within spin- and angle- resolved photoemission spectroscopy separating the in-plane orbitals ($p_x$/$p_y$) from the out-of-plane ($p_z$).\cite{NATUREXie2014}} 

In order to achieve controllable parameters in this system, we extend our discussion to a bilayer structure to explore the interlayer coupling between C1-C1 and C1-C2 stackings, and the effects of a perpendicular electric field. 

\subsection{Control of orbital-texture in coupled bilayer}

\begin{figure}
\includegraphics[width=\columnwidth]{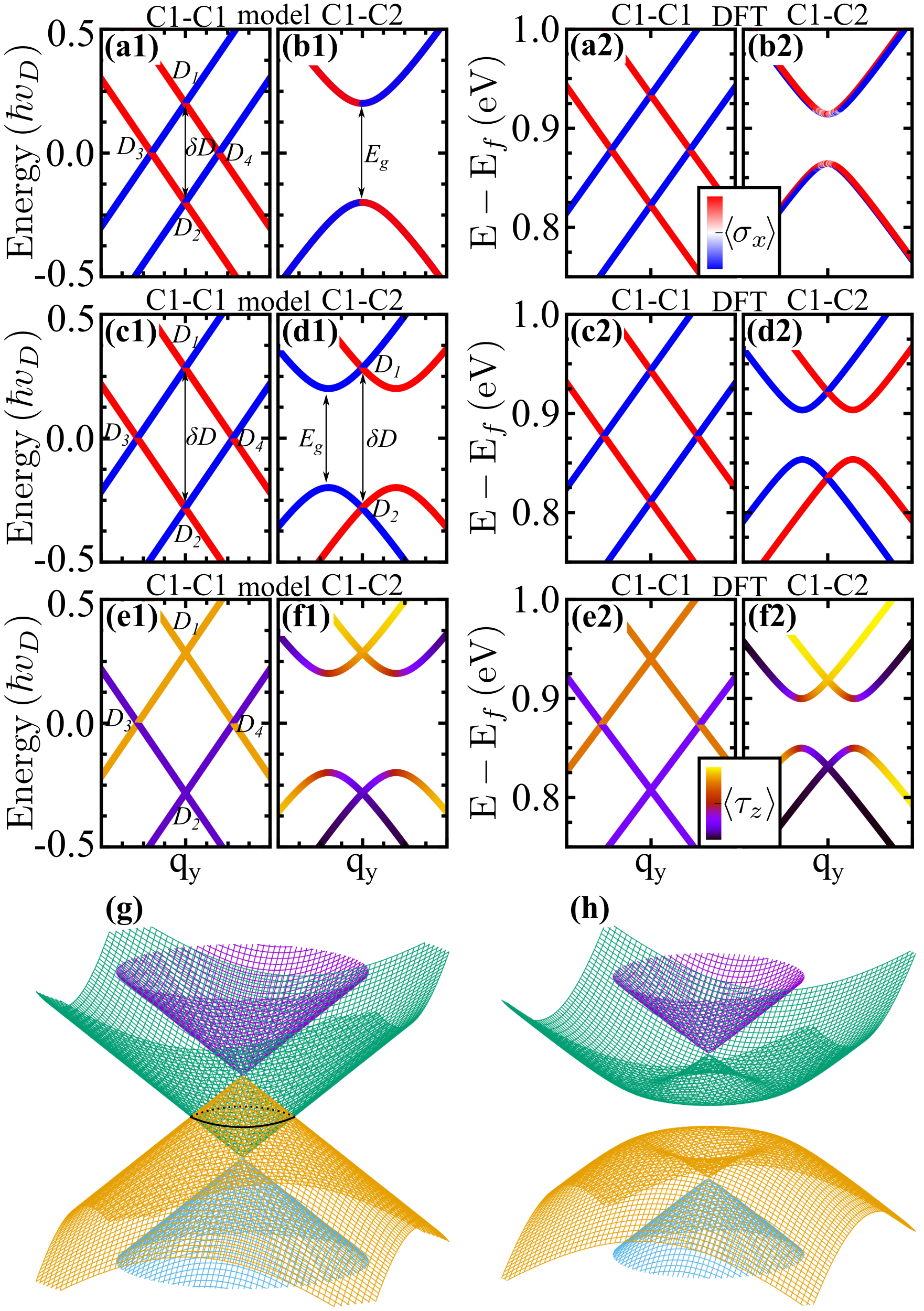}
\caption{\label{bilayer} Bilayer model orbital-texture projected along {$\vec{q} =(0,\,q_y)$} dispersion (a1)-(d1), and DFT orbital-texture (a2)-(d2). The zero electric field and same chirality stacking is show in (a1) and (a2); while for opposite chirality stacking is (b1) and (b2). The finite electric field and same chirality stacking is show in (c1) and (c2); while for opposite chirality stacking is (d1) and (d2). Layer pseudospin polarization ($\avg{\tau_z}$) projected dispersion with external electric field for same chirality stacking is show in (e1) for the model and (e2) for DFT; while for opposite chirality stacking is (f1) for the model and (f2) for DFT. 3D band structure plot for finite electric field for the same (g) and opposite (h) chirality stacking.}
\end{figure}

First, let us discuss the extension of our model for the stacking of monolayers of equal chirality, which we label C1-C1 stacking, and investigate the consequences of the interlayer coupling the orbital texture. Later, as a proof of principle, we compare the analytic results with a DFT simulated bilayer stacking, {where in each case we take the most symmetric stacking, i.e. with aligned hexagonal vacancy centers}. Since our basis $\ket{\pm}$ is composed by planar $(p_x,p_y)$ orbitals, and the stacking does not break in-plane symmetries, the interlayer coupling $\Delta$ must couple only states with equivalent symmetries on each layer \cite{JCPCrasto2019}. Additionally, we consider an out-of-plane external electric field by changing the on-site energy of each layer. Thus, at the $K$ point, the C1-C1 stacking hamiltonian is
\begin{equation}
H^{(11)}_K = h' \otimes \tau_0 + \Delta\, \mathbb{I} \otimes \tau_x + \nu\, \mathbb{I} \otimes \tau_z. \label{eq:Hs}
\end{equation}
Here, $h'={\rm diag}(e_+,\,e_-)$ is the monolayer hamiltonian in its diagonal base $\left\{\ket{e_\lambda} \right\}$, $\mathbb{I}$ is a $2\times 2$ identity matrix, $\tau_x$, $\tau_y$, and $\tau_z$ are the Pauli matrices in the layer subspace, $\Delta$ is the interlayer coupling strength, and $\nu$ the potential energy in the layers due to the out-of-plane electric field. The eigenvalues are given by
\begin{equation}
E_{m,\,\lambda}^{(11)} = e_{\lambda} + m\,\sqrt{\Delta^2 + \nu^2},\label{eq:eig-Hs}
\end{equation}
where $e_{\lambda}$ is the monolayer energy dispersion, $m$ and $\lambda=\pm1$ are, respectively, the quantum numbers associated with mirror (for $\nu=0$) and helicity symmetries, that define the eigenvectors
\begin{equation}
\ket{m,\,e_\lambda} = \frac{\ket{e_\lambda} \otimes \left( \ket{b} + g_m \ket{t} \right)}{\sqrt{1+g_m^2}}.\label{eq:eiv-Hs}
\end{equation}
Here, the $m$ dependent factor $g_m = \Delta/(\nu + m\sqrt{\Delta^2 + \nu^2})$, and the kets $\ket{b}$ and $\ket{t}$ represents the states of the bottom and top layers, respectively. Note that the Hamiltonian of each layer is the same ($h'$), due to their equivalent chirality. Therefore, for zero external field $\nu=0$, the Hamiltonian commutes with $\tau_x$ (which coincide with the $M_z$ symmetry operator), and the eigenstates reduce to the mirror symmetric and antisymmetric cases $g_m=\pm1$. {Moreover, for $\nu \neq 0$ we can still separate the Hamiltonian into two orthogonal subspaces defined by its pseudomirror symmetry\cite{PRBdeLima2017} $\tilde{M}_z = (\Delta \tau_x + \nu \tau_z)/\sqrt{\Delta^2 + \nu^2}$}. Mirror symmetry aside, the $\Delta$ and $\nu$ parameters enter in the energy dispersion in the same way, separating the two Dirac cones ($D_1$ and $D_2$) in energy by $\delta D = 2\sqrt{\Delta^2 + \nu^2}$, see Fig.\,\ref{bilayer}(a1) for $\nu=0$ and (c1) for $\nu \neq 0$.

Indeed, the DFT calculations\cite{note1} including an external electric field of $E^{\rm ext}=0.02$\,eV/{\AA} [Fig.\,\ref{bilayer}(c2)] show an increase in the separation of the two Dirac crossing ($\delta D$), $0.111 \rightarrow 0.132$\,eV, compared with the $E^{\rm ext}=0$\,eV/{\AA} case [Fig.\,\ref{bilayer}(a2)]. Such separation of the Dirac cones lead to a linear dispersive nodal line, see Fig.\,\ref{bilayer}(g), with the crossing originated between bands with opposite mirror eigenvalues [$D_3$ and $D_4$, in Fig.\,\ref{bilayer}(a1) and (c1)]. Furthermore, as the orbital texture depends on $\ket{\lambda}$ only, in this stacking it becomes independent of $m$,
\begin{equation}
\bra{m,\,e_\lambda}\sigma_i \ket{m,\,e_\lambda} = \mean{\sigma_i}_\lambda,\;\;\;\,\, i=x,\,y.\label{eq:txt-Hs}
\end{equation} 
Therefore for each energy split Dirac cones ($D_1$ e $D_2$), the orbital texture will keep the same behavior as in the monolayer. In this C1-C1 stacking, the electric field affects only the layer degree of freedom. Where for zero field ($\nu=0$) we have the symmetric and anti-symmetric states with same contribution of each layer, i.e. $|\avg{b|m}|^2 = |\avg{t|m}|^2=0.5$. On the other hand, including the electric field the states start to localize in one layer or the other, see the color code in Fig.\,\ref{bilayer}(e1) and (e2). Furthermore, the decoupling of the orbital texture from the layer degree of freedom, as shown in Eq.\,\eqref{eq:eiv-Hs}, allows interesting control in the $D_3$ and $D_4$ crossing. Here, the orbital-texture is robust against electric field changes, but the layer-pseudo spin changes from a unpolarized ($\avg{\tau_z}=0$) crossing for $\nu=0$, to a crossing of states mostly localized in different layers ($\avg{\tau_z}\neq 0$), see Fig.\,\ref{bilayer}(e1).

A distinct behavior will arise for stackings of layers with opposite chiralities, C1-C2 stacking. The Hamiltonian in this case is built exchanging the top layer for an opposite chirality monolayer, for which its monolayer hamiltonian, $h$, changes sign. Therefore the hamiltonian is written as in the same chirality case, Eq.\,\eqref{eq:Hs}, but replacing $\tau_0$ for $\tau_z$ in the fist term. The eigenvalues are given by
\begin{equation}
E_{m,\,\lambda}^{(12)} = m \sqrt{\left( e_\lambda + \nu \right)^2 + \Delta^2},\label{eq:eig-Hd}
\end{equation}
with $e_\lambda$ being the monolayer dispersion, and $m$ and $\lambda=\pm1$. Correspondingly, the eigenvectors are 
{
\begin{equation}
\ket{m,\,e_\lambda} = \frac{ \ket{e_\lambda} \otimes \left( \ket{b} + f_{m,\,\lambda} \ket{t} \right)}{\sqrt{1+f_{m,\,\lambda}^2}}. \label{eq:eiv-Hd}
\end{equation}
}
where $f_{m,\,\lambda}$ is equivalent to $g_m$ with the replecement $\nu \rightarrow \left( e_\lambda + \nu \right)$. {Note that for the C1-C2 stacking the orthogonality constrain of the C1-C1 case is not present, as the Hamiltonian does not commute with the mirror $M_z$ nor pseudomirror $\tilde{M}_z$ operators}. Here a parabolic type of dispersion takes place close to the $K$ point for $\nu=0$, Fig.\,\ref{bilayer}(b1), with an energy gap of $E_g=2\Delta$. By turning on the electric field $\nu \neq 0$, a recovery of the $D_1$ and $D_2$ Dirac crossing is observed, see Fig.\,\ref{bilayer}(d1), with its energy separation being the same as in the C1-C1 case, while the energy gap $E_g$ remains the same as the $\nu=0$ case. {Note that the mechanism responsible for the gap opening, i.e. the interlayer coupling, is unaltered by the presence of the electric field, the latter only dislocate the degeneracy point.}The same behavior is observed for the DFT results with or without an external field of $E^{\rm ext}=0.02$\,eV/{\AA}, Fig.\,\ref{bilayer}(b2) and (d2). {In this bilayer C1-C2 stacking, the eigenstates separation of layer and orbital degrees of freedom lead to the same expression for the orbital texture as in Eq.\,\eqref{eq:txt-Hs}. Such texture is indeed corroborated by the {\fp} calculations, Fig.\,\ref{bilayer}(d1) and (d2). Here, at the $D_1$ and $D_2$ crossing, the helical orbital-texture is preserved, but with opposite crossings.} Additionally, the electric field also change the layer contribution to each state, with a localization effect of the $D_1$ and $D_2$ cones in the bottom and top layer respectively, see Fig.\,\ref{bilayer}(f1) and (f2). Despite the preservation of $D_1$ and $D_2$ crossings in the presence of electric fields, by changing the stacking from same to opposite chirality, the Dirac nodal line opens a gap, compare Fig.\,\ref{bilayer}(g) and (h).

Furthermore, an analogy with the TI surface states can be made for the C1-C2 stacking. Note that each layer (top and bottom) has opposite orbital texture, as the opposite spin texture in each TI surface state. In an thin film confinement of 3D topological insulator with an external electric field, a coupling and asymmetry between each opposite surface arises. The combinations of this two effects lead to a surface localization of the topological states in a similar way as the layer localization observed in the orbital textures \cite{NJPShan2010}.

\section{Conclusion}

We have shown the existence of orbital-texture in a 2D chiral borophene monolayer phase in the absence of SOC. This texture arises in a Dirac cone formed by a $p_x/p_y$ orbitals, defining a pseudospin  with its direction locked with the momentum, and with opposed textures at K and K' valleys. The high Dirac velocity found in this system indicates a {possible} enhancement of the orbital Edelstein effect. Extending our model for a borophene bilayers with layers of equal (C1-C1) or opposite (C1-C2) chiralities, we find a possible external control of the orbital textures. In the former case we found a conservation of the monolayer orbital texture, and a Dirac node line {with a radius defined by the electric field ($\nu$) and interlayer coupling strength ($\Delta$)}. In the opposite chirality stacking (C1-C2), the interplay of orbital-texture with the layer degree of freedom brings another feature to future applications. For instance, in this stacking, the electric field induces a controllable {on/off} switch of the Dirac cones at the K/K' point and, consequently, the orbital-textures, which becomes localized on different layers.

\section{ACKNOWLEDGMENTS}

The authors acknowledge   financial   support   from   the Brazilian  agencies  
CNPq, and FAPEMIG, and the CENAPAD-SP and Laborat\'{o}rio Nacional de Computa\c{c}\~{a}o Cient\'{i}fica (LNCC-SCAFMat) for computer time.

\appendix

\section{Computational approach} \label{app:met}
The calculations of atomic geometry and band structure were performed based on the DFT approach, as implemented in the VASP code \cite{VASP}. The exchange correlation term was described using the GGA functional proposed by Perdew, Burke, and Ernzerhof \cite{PBE}. The Kohn-Sham orbitals are expanded in a plane wave basis set with an energy cutoff of $450$\,eV. The Brillouin zone is sampled according to the Monkhorst-Pack method \cite{PhysRevB.13.5188}, using a gamma-centered $7 \times 7 \times 1$ mesh. The electron-ion interactions are taken into account using the Projector Augmented Wave method \cite{PAW}. All geometries have been relaxed until atomic forces were lower than 0.01 eV/{\AA}.

\bibliography{bib}

\begin{thebibliography}{47}%
\makeatletter
\providecommand \@ifxundefined [1]{%
 \@ifx{#1\undefined}
}%
\providecommand \@ifnum [1]{%
 \ifnum #1\expandafter \@firstoftwo
 \else \expandafter \@secondoftwo
 \fi
}%
\providecommand \@ifx [1]{%
 \ifx #1\expandafter \@firstoftwo
 \else \expandafter \@secondoftwo
 \fi
}%
\providecommand \natexlab [1]{#1}%
\providecommand \enquote  [1]{``#1''}%
\providecommand \bibnamefont  [1]{#1}%
\providecommand \bibfnamefont [1]{#1}%
\providecommand \citenamefont [1]{#1}%
\providecommand \href@noop [0]{\@secondoftwo}%
\providecommand \href [0]{\begingroup \@sanitize@url \@href}%
\providecommand \@href[1]{\@@startlink{#1}\@@href}%
\providecommand \@@href[1]{\endgroup#1\@@endlink}%
\providecommand \@sanitize@url [0]{\catcode `\\12\catcode `\$12\catcode
  `\&12\catcode `\#12\catcode `\^12\catcode `\_12\catcode `\%12\relax}%
\providecommand \@@startlink[1]{}%
\providecommand \@@endlink[0]{}%
\providecommand \url  [0]{\begingroup\@sanitize@url \@url }%
\providecommand \@url [1]{\endgroup\@href {#1}{\urlprefix }}%
\providecommand \urlprefix  [0]{URL }%
\providecommand \Eprint [0]{\href }%
\providecommand \doibase [0]{http://dx.doi.org/}%
\providecommand \selectlanguage [0]{\@gobble}%
\providecommand \bibinfo  [0]{\@secondoftwo}%
\providecommand \bibfield  [0]{\@secondoftwo}%
\providecommand \translation [1]{[#1]}%
\providecommand \BibitemOpen [0]{}%
\providecommand \bibitemStop [0]{}%
\providecommand \bibitemNoStop [0]{.\EOS\space}%
\providecommand \EOS [0]{\spacefactor3000\relax}%
\providecommand \BibitemShut  [1]{\csname bibitem#1\endcsname}%
\let\auto@bib@innerbib\@empty
\bibitem [{\citenamefont {Novoselov}\ \emph {et~al.}(2004)\citenamefont
  {Novoselov}, \citenamefont {Geim}, \citenamefont {Morozov}, \citenamefont
  {Jiang}, \citenamefont {Zhang}, \citenamefont {Dubonos}, \citenamefont
  {Grigorieva},\ and\ \citenamefont {Firsov}}]{SCIENCENovoselov2004}%
  \BibitemOpen
  \bibfield  {author} {\bibinfo {author} {\bibfnamefont {K.~S.}\ \bibnamefont
  {Novoselov}}, \bibinfo {author} {\bibfnamefont {A.~K.}\ \bibnamefont {Geim}},
  \bibinfo {author} {\bibfnamefont {S.~V.}\ \bibnamefont {Morozov}}, \bibinfo
  {author} {\bibfnamefont {D.}~\bibnamefont {Jiang}}, \bibinfo {author}
  {\bibfnamefont {Y.}~\bibnamefont {Zhang}}, \bibinfo {author} {\bibfnamefont
  {S.~V.}\ \bibnamefont {Dubonos}}, \bibinfo {author} {\bibfnamefont {I.~V.}\
  \bibnamefont {Grigorieva}}, \ and\ \bibinfo {author} {\bibfnamefont {A.~A.}\
  \bibnamefont {Firsov}},\ }\bibfield  {title} {\enquote {\bibinfo {title}
  {Electric field effect in atomically thin carbon films},}\ }\href {\doibase
  10.1126/science.1102896} {\bibfield  {journal} {\bibinfo  {journal}
  {Science}\ }\textbf {\bibinfo {volume} {306}},\ \bibinfo {pages} {666--669}
  (\bibinfo {year} {2004})}\BibitemShut {NoStop}%
\bibitem [{\citenamefont {Mas-Ballest\'{e}}\ \emph {et~al.}(2011)\citenamefont
  {Mas-Ballest\'{e}}, \citenamefont {G\'{o}mez-Navarro}, \citenamefont
  {G\'{o}mez-Herrero},\ and\ \citenamefont {Zamora}}]{NANOSCALEMas2011}%
  \BibitemOpen
  \bibfield  {author} {\bibinfo {author} {\bibfnamefont {Rub\'{e}n}\
  \bibnamefont {Mas-Ballest\'{e}}}, \bibinfo {author} {\bibfnamefont
  {Cristina}\ \bibnamefont {G\'{o}mez-Navarro}}, \bibinfo {author}
  {\bibfnamefont {Julio}\ \bibnamefont {G\'{o}mez-Herrero}}, \ and\ \bibinfo
  {author} {\bibfnamefont {F\'{e}lix}\ \bibnamefont {Zamora}},\ }\bibfield
  {title} {\enquote {\bibinfo {title} {2d materials: to graphene and beyond},}\
  }\href {\doibase 10.1039/C0NR00323A} {\bibfield  {journal} {\bibinfo
  {journal} {Nanoscale}\ }\textbf {\bibinfo {volume} {3}},\ \bibinfo {pages}
  {20--30} (\bibinfo {year} {2011})}\BibitemShut {NoStop}%
\bibitem [{\citenamefont {Novoselov}\ \emph {et~al.}(2016)\citenamefont
  {Novoselov}, \citenamefont {Mishchenko}, \citenamefont {Carvalho},\ and\
  \citenamefont {Castro~Neto}}]{SCIENCENovoselov2016}%
  \BibitemOpen
  \bibfield  {author} {\bibinfo {author} {\bibfnamefont {K.~S.}\ \bibnamefont
  {Novoselov}}, \bibinfo {author} {\bibfnamefont {A.}~\bibnamefont
  {Mishchenko}}, \bibinfo {author} {\bibfnamefont {A.}~\bibnamefont
  {Carvalho}}, \ and\ \bibinfo {author} {\bibfnamefont {A.~H.}\ \bibnamefont
  {Castro~Neto}},\ }\bibfield  {title} {\enquote {\bibinfo {title} {2d
  materials and van der waals heterostructures},}\ }\href {\doibase
  10.1126/science.aac9439} {\bibfield  {journal} {\bibinfo  {journal}
  {Science}\ }\textbf {\bibinfo {volume} {353}} (\bibinfo {year} {2016}),\
  10.1126/science.aac9439}\BibitemShut {NoStop}%
\bibitem [{\citenamefont {Castro~Neto}\ \emph {et~al.}(2009)\citenamefont
  {Castro~Neto}, \citenamefont {Guinea}, \citenamefont {Peres}, \citenamefont
  {Novoselov},\ and\ \citenamefont {Geim}}]{RMPCastro2009}%
  \BibitemOpen
  \bibfield  {author} {\bibinfo {author} {\bibfnamefont {A.~H.}\ \bibnamefont
  {Castro~Neto}}, \bibinfo {author} {\bibfnamefont {F.}~\bibnamefont {Guinea}},
  \bibinfo {author} {\bibfnamefont {N.~M.~R.}\ \bibnamefont {Peres}}, \bibinfo
  {author} {\bibfnamefont {K.~S.}\ \bibnamefont {Novoselov}}, \ and\ \bibinfo
  {author} {\bibfnamefont {A.~K.}\ \bibnamefont {Geim}},\ }\bibfield  {title}
  {\enquote {\bibinfo {title} {The electronic properties of graphene},}\ }\href
  {\doibase 10.1103/RevModPhys.81.109} {\bibfield  {journal} {\bibinfo
  {journal} {Rev. Mod. Phys.}\ }\textbf {\bibinfo {volume} {81}},\ \bibinfo
  {pages} {109--162} (\bibinfo {year} {2009})}\BibitemShut {NoStop}%
\bibitem [{\citenamefont {Fu}\ \emph {et~al.}(2007)\citenamefont {Fu},
  \citenamefont {Kane},\ and\ \citenamefont {Mele}}]{PRLFu2007}%
  \BibitemOpen
  \bibfield  {author} {\bibinfo {author} {\bibfnamefont {Liang}\ \bibnamefont
  {Fu}}, \bibinfo {author} {\bibfnamefont {C.~L.}\ \bibnamefont {Kane}}, \ and\
  \bibinfo {author} {\bibfnamefont {E.~J.}\ \bibnamefont {Mele}},\ }\bibfield
  {title} {\enquote {\bibinfo {title} {Topological insulators in three
  dimensions},}\ }\href {\doibase 10.1103/PhysRevLett.98.106803} {\bibfield
  {journal} {\bibinfo  {journal} {Phys. Rev. Lett.}\ }\textbf {\bibinfo
  {volume} {98}},\ \bibinfo {pages} {106803} (\bibinfo {year}
  {2007})}\BibitemShut {NoStop}%
\bibitem [{\citenamefont {Hasan}\ and\ \citenamefont
  {Kane}(2010)}]{RMPHasan2010}%
  \BibitemOpen
  \bibfield  {author} {\bibinfo {author} {\bibfnamefont {M.~Z.}\ \bibnamefont
  {Hasan}}\ and\ \bibinfo {author} {\bibfnamefont {C.~L.}\ \bibnamefont
  {Kane}},\ }\bibfield  {title} {\enquote {\bibinfo {title} {Colloquium:
  Topological insulators},}\ }\href {\doibase 10.1103/RevModPhys.82.3045}
  {\bibfield  {journal} {\bibinfo  {journal} {Rev. Mod. Phys.}\ }\textbf
  {\bibinfo {volume} {82}},\ \bibinfo {pages} {3045--3067} (\bibinfo {year}
  {2010})}\BibitemShut {NoStop}%
\bibitem [{\citenamefont {Edelstein}(1990)}]{SSCEdelstein1990}%
  \BibitemOpen
  \bibfield  {author} {\bibinfo {author} {\bibfnamefont {V.M.}\ \bibnamefont
  {Edelstein}},\ }\bibfield  {title} {\enquote {\bibinfo {title} {Spin
  polarization of conduction electrons induced by electric current in
  two-dimensional asymmetric electron systems},}\ }\href {\doibase
  https://doi.org/10.1016/0038-1098(90)90963-C} {\bibfield  {journal} {\bibinfo
   {journal} {Solid State Communications}\ }\textbf {\bibinfo {volume} {73}},\
  \bibinfo {pages} {233 -- 235} (\bibinfo {year} {1990})}\BibitemShut {NoStop}%
\bibitem [{\citenamefont {Silsbee}(2001)}]{PRBSilsbee2001}%
  \BibitemOpen
  \bibfield  {author} {\bibinfo {author} {\bibfnamefont {Robert~H.}\
  \bibnamefont {Silsbee}},\ }\bibfield  {title} {\enquote {\bibinfo {title}
  {Theory of the detection of current-induced spin polarization in a
  two-dimensional electron gas},}\ }\href {\doibase 10.1103/PhysRevB.63.155305}
  {\bibfield  {journal} {\bibinfo  {journal} {Phys. Rev. B}\ }\textbf {\bibinfo
  {volume} {63}},\ \bibinfo {pages} {155305} (\bibinfo {year}
  {2001})}\BibitemShut {NoStop}%
\bibitem [{\citenamefont {Rodriguez-Vega}\ \emph {et~al.}(2017)\citenamefont
  {Rodriguez-Vega}, \citenamefont {Schwiete}, \citenamefont {Sinova},\ and\
  \citenamefont {Rossi}}]{PRBRodriguez2017}%
  \BibitemOpen
  \bibfield  {author} {\bibinfo {author} {\bibfnamefont {M.}~\bibnamefont
  {Rodriguez-Vega}}, \bibinfo {author} {\bibfnamefont {G.}~\bibnamefont
  {Schwiete}}, \bibinfo {author} {\bibfnamefont {J.}~\bibnamefont {Sinova}}, \
  and\ \bibinfo {author} {\bibfnamefont {E.}~\bibnamefont {Rossi}},\ }\bibfield
   {title} {\enquote {\bibinfo {title} {Giant edelstein effect in
  topological-insulator--graphene heterostructures},}\ }\href {\doibase
  10.1103/PhysRevB.96.235419} {\bibfield  {journal} {\bibinfo  {journal} {Phys.
  Rev. B}\ }\textbf {\bibinfo {volume} {96}},\ \bibinfo {pages} {235419}
  (\bibinfo {year} {2017})}\BibitemShut {NoStop}%
\bibitem [{\citenamefont {Geng}\ \emph {et~al.}(2017)\citenamefont {Geng},
  \citenamefont {Luo}, \citenamefont {Deng}, \citenamefont {Sheng},
  \citenamefont {Shen},\ and\ \citenamefont {Xing}}]{SCIREPGeng2017}%
  \BibitemOpen
  \bibfield  {author} {\bibinfo {author} {\bibfnamefont {H.}~\bibnamefont
  {Geng}}, \bibinfo {author} {\bibfnamefont {W.}~\bibnamefont {Luo}}, \bibinfo
  {author} {\bibfnamefont {W.~Y.}\ \bibnamefont {Deng}}, \bibinfo {author}
  {\bibfnamefont {L.}~\bibnamefont {Sheng}}, \bibinfo {author} {\bibfnamefont
  {R.}~\bibnamefont {Shen}}, \ and\ \bibinfo {author} {\bibfnamefont {D.~Y.}\
  \bibnamefont {Xing}},\ }\bibfield  {title} {\enquote {\bibinfo {title}
  {Theory of inverse edelstein effect of the surface states of a topological
  insulator},}\ }\href {\doibase 10.1038/s41598-017-03346-z} {\bibfield
  {journal} {\bibinfo  {journal} {Scientific Reports}\ }\textbf {\bibinfo
  {volume} {7}},\ \bibinfo {pages} {3755} (\bibinfo {year} {2017})}\BibitemShut
  {NoStop}%
\bibitem [{\citenamefont {Zhang}\ \emph {et~al.}(2013)\citenamefont {Zhang},
  \citenamefont {Liu},\ and\ \citenamefont {Zhang}}]{PRLZhang2013}%
  \BibitemOpen
  \bibfield  {author} {\bibinfo {author} {\bibfnamefont {Haijun}\ \bibnamefont
  {Zhang}}, \bibinfo {author} {\bibfnamefont {Chao-Xing}\ \bibnamefont {Liu}},
  \ and\ \bibinfo {author} {\bibfnamefont {Shou-Cheng}\ \bibnamefont {Zhang}},\
  }\bibfield  {title} {\enquote {\bibinfo {title} {Spin-orbital texture in
  topological insulators},}\ }\href {\doibase 10.1103/PhysRevLett.111.066801}
  {\bibfield  {journal} {\bibinfo  {journal} {Phys. Rev. Lett.}\ }\textbf
  {\bibinfo {volume} {111}},\ \bibinfo {pages} {066801} (\bibinfo {year}
  {2013})}\BibitemShut {NoStop}%
\bibitem [{\citenamefont {Waugh}\ \emph {et~al.}(2016)\citenamefont {Waugh},
  \citenamefont {Nummy}, \citenamefont {Parham}, \citenamefont {Liu},
  \citenamefont {Zhang}, \citenamefont {Zunger},\ and\ \citenamefont
  {Dessau}}]{NATUREWaugh2016}%
  \BibitemOpen
  \bibfield  {author} {\bibinfo {author} {\bibfnamefont {Justin~A.}\
  \bibnamefont {Waugh}}, \bibinfo {author} {\bibfnamefont {Thomas}\
  \bibnamefont {Nummy}}, \bibinfo {author} {\bibfnamefont {Stephen}\
  \bibnamefont {Parham}}, \bibinfo {author} {\bibfnamefont {Qihang}\
  \bibnamefont {Liu}}, \bibinfo {author} {\bibfnamefont {Xiuwen}\ \bibnamefont
  {Zhang}}, \bibinfo {author} {\bibfnamefont {Alex}\ \bibnamefont {Zunger}}, \
  and\ \bibinfo {author} {\bibfnamefont {Daniel~S.}\ \bibnamefont {Dessau}},\
  }\bibfield  {title} {\enquote {\bibinfo {title} {Minimal ingredients for
  orbital-texture switches at dirac points in strong spin-orbit coupled
  materials},}\ }\href {https://doi.org/10.1038/npjquantmats.2016.25}
  {\bibfield  {journal} {\bibinfo  {journal} {Npj Quantum Materials}\ }\textbf
  {\bibinfo {volume} {1}},\ \bibinfo {pages} {16025} (\bibinfo {year}
  {2016})}\BibitemShut {NoStop}%
\bibitem [{\citenamefont {Gotlieb}\ \emph {et~al.}(2017)\citenamefont
  {Gotlieb}, \citenamefont {Li}, \citenamefont {Lin}, \citenamefont {Jozwiak},
  \citenamefont {Ryoo}, \citenamefont {Park}, \citenamefont {Hussain},
  \citenamefont {Louie},\ and\ \citenamefont {Lanzara}}]{PRBGotlieb2017}%
  \BibitemOpen
  \bibfield  {author} {\bibinfo {author} {\bibfnamefont {Kenneth}\ \bibnamefont
  {Gotlieb}}, \bibinfo {author} {\bibfnamefont {Zhenglu}\ \bibnamefont {Li}},
  \bibinfo {author} {\bibfnamefont {Chiu-Yun}\ \bibnamefont {Lin}}, \bibinfo
  {author} {\bibfnamefont {Chris}\ \bibnamefont {Jozwiak}}, \bibinfo {author}
  {\bibfnamefont {Ji~Hoon}\ \bibnamefont {Ryoo}}, \bibinfo {author}
  {\bibfnamefont {Cheol-Hwan}\ \bibnamefont {Park}}, \bibinfo {author}
  {\bibfnamefont {Zahid}\ \bibnamefont {Hussain}}, \bibinfo {author}
  {\bibfnamefont {Steven~G.}\ \bibnamefont {Louie}}, \ and\ \bibinfo {author}
  {\bibfnamefont {Alessandra}\ \bibnamefont {Lanzara}},\ }\bibfield  {title}
  {\enquote {\bibinfo {title} {Symmetry rules shaping spin-orbital textures in
  surface states},}\ }\href {\doibase 10.1103/PhysRevB.95.245142} {\bibfield
  {journal} {\bibinfo  {journal} {Phys. Rev. B}\ }\textbf {\bibinfo {volume}
  {95}},\ \bibinfo {pages} {245142} (\bibinfo {year} {2017})}\BibitemShut
  {NoStop}%
\bibitem [{\citenamefont {Xie}\ \emph {et~al.}(2014)\citenamefont {Xie},
  \citenamefont {He}, \citenamefont {Chen}, \citenamefont {Feng}, \citenamefont
  {Yi}, \citenamefont {Liang}, \citenamefont {Zhao}, \citenamefont {Mou},
  \citenamefont {He}, \citenamefont {Peng}, \citenamefont {Liu}, \citenamefont
  {Liu}, \citenamefont {Liu}, \citenamefont {Dong}, \citenamefont {Yu},
  \citenamefont {Zhang}, \citenamefont {Zhang}, \citenamefont {Wang},
  \citenamefont {Zhang}, \citenamefont {Yang}, \citenamefont {Peng},
  \citenamefont {Wang}, \citenamefont {Chen}, \citenamefont {Xu},\ and\
  \citenamefont {Zhou}}]{NATUREXie2014}%
  \BibitemOpen
  \bibfield  {author} {\bibinfo {author} {\bibfnamefont {Zhuojin}\ \bibnamefont
  {Xie}}, \bibinfo {author} {\bibfnamefont {Shaolong}\ \bibnamefont {He}},
  \bibinfo {author} {\bibfnamefont {Chaoyu}\ \bibnamefont {Chen}}, \bibinfo
  {author} {\bibfnamefont {Ya}~\bibnamefont {Feng}}, \bibinfo {author}
  {\bibfnamefont {Hemian}\ \bibnamefont {Yi}}, \bibinfo {author} {\bibfnamefont
  {Aiji}\ \bibnamefont {Liang}}, \bibinfo {author} {\bibfnamefont {Lin}\
  \bibnamefont {Zhao}}, \bibinfo {author} {\bibfnamefont {Daixiang}\
  \bibnamefont {Mou}}, \bibinfo {author} {\bibfnamefont {Junfeng}\ \bibnamefont
  {He}}, \bibinfo {author} {\bibfnamefont {Yingying}\ \bibnamefont {Peng}},
  \bibinfo {author} {\bibfnamefont {Xu}~\bibnamefont {Liu}}, \bibinfo {author}
  {\bibfnamefont {Yan}\ \bibnamefont {Liu}}, \bibinfo {author} {\bibfnamefont
  {Guodong}\ \bibnamefont {Liu}}, \bibinfo {author} {\bibfnamefont {Xiaoli}\
  \bibnamefont {Dong}}, \bibinfo {author} {\bibfnamefont {Li}~\bibnamefont
  {Yu}}, \bibinfo {author} {\bibfnamefont {Jun}\ \bibnamefont {Zhang}},
  \bibinfo {author} {\bibfnamefont {Shenjin}\ \bibnamefont {Zhang}}, \bibinfo
  {author} {\bibfnamefont {Zhimin}\ \bibnamefont {Wang}}, \bibinfo {author}
  {\bibfnamefont {Fengfeng}\ \bibnamefont {Zhang}}, \bibinfo {author}
  {\bibfnamefont {Feng}\ \bibnamefont {Yang}}, \bibinfo {author} {\bibfnamefont
  {Qinjun}\ \bibnamefont {Peng}}, \bibinfo {author} {\bibfnamefont {Xiaoyang}\
  \bibnamefont {Wang}}, \bibinfo {author} {\bibfnamefont {Chuangtian}\
  \bibnamefont {Chen}}, \bibinfo {author} {\bibfnamefont {Zuyan}\ \bibnamefont
  {Xu}}, \ and\ \bibinfo {author} {\bibfnamefont {X.~J.}\ \bibnamefont
  {Zhou}},\ }\bibfield  {title} {\enquote {\bibinfo {title} {Orbital-selective
  spin texture and its manipulation in a topological insulator},}\ }\href
  {https://doi.org/10.1038/ncomms4382} {\bibfield  {journal} {\bibinfo
  {journal} {Nature Communications}\ }\textbf {\bibinfo {volume} {5}},\
  \bibinfo {pages} {3382} (\bibinfo {year} {2014})}\BibitemShut {NoStop}%
\bibitem [{\citenamefont {Miao}\ \emph {et~al.}(2014)\citenamefont {Miao},
  \citenamefont {Wang}, \citenamefont {Yao}, \citenamefont {Zhu}, \citenamefont
  {Dil}, \citenamefont {Gao}, \citenamefont {Liu}, \citenamefont {Liu},
  \citenamefont {Qian},\ and\ \citenamefont {Jia}}]{PRBMiao2014}%
  \BibitemOpen
  \bibfield  {author} {\bibinfo {author} {\bibfnamefont {Lin}\ \bibnamefont
  {Miao}}, \bibinfo {author} {\bibfnamefont {Z.~F.}\ \bibnamefont {Wang}},
  \bibinfo {author} {\bibfnamefont {Meng-Yu}\ \bibnamefont {Yao}}, \bibinfo
  {author} {\bibfnamefont {Fengfeng}\ \bibnamefont {Zhu}}, \bibinfo {author}
  {\bibfnamefont {J.~H.}\ \bibnamefont {Dil}}, \bibinfo {author} {\bibfnamefont
  {C.~L.}\ \bibnamefont {Gao}}, \bibinfo {author} {\bibfnamefont {Canhua}\
  \bibnamefont {Liu}}, \bibinfo {author} {\bibfnamefont {Feng}\ \bibnamefont
  {Liu}}, \bibinfo {author} {\bibfnamefont {Dong}\ \bibnamefont {Qian}}, \ and\
  \bibinfo {author} {\bibfnamefont {Jin-Feng}\ \bibnamefont {Jia}},\ }\bibfield
   {title} {\enquote {\bibinfo {title} {Orbit- and atom-resolved spin textures
  of intrinsic, extrinsic, and hybridized dirac cone states},}\ }\href
  {\doibase 10.1103/PhysRevB.89.155116} {\bibfield  {journal} {\bibinfo
  {journal} {Phys. Rev. B}\ }\textbf {\bibinfo {volume} {89}},\ \bibinfo
  {pages} {155116} (\bibinfo {year} {2014})}\BibitemShut {NoStop}%
\bibitem [{\citenamefont {Kontani}\ \emph {et~al.}(2009)\citenamefont
  {Kontani}, \citenamefont {Tanaka}, \citenamefont {Hirashima}, \citenamefont
  {Yamada},\ and\ \citenamefont {Inoue}}]{PRLGiant2009}%
  \BibitemOpen
  \bibfield  {author} {\bibinfo {author} {\bibfnamefont {H.}~\bibnamefont
  {Kontani}}, \bibinfo {author} {\bibfnamefont {T.}~\bibnamefont {Tanaka}},
  \bibinfo {author} {\bibfnamefont {D.~S.}\ \bibnamefont {Hirashima}}, \bibinfo
  {author} {\bibfnamefont {K.}~\bibnamefont {Yamada}}, \ and\ \bibinfo {author}
  {\bibfnamefont {J.}~\bibnamefont {Inoue}},\ }\bibfield  {title} {\enquote
  {\bibinfo {title} {Giant orbital hall effect in transition metals: Origin of
  large spin and anomalous hall effects},}\ }\href {\doibase
  10.1103/PhysRevLett.102.016601} {\bibfield  {journal} {\bibinfo  {journal}
  {Phys. Rev. Lett.}\ }\textbf {\bibinfo {volume} {102}},\ \bibinfo {pages}
  {016601} (\bibinfo {year} {2009})}\BibitemShut {NoStop}%
\bibitem [{\citenamefont {Jo}\ \emph {et~al.}(2018)\citenamefont {Jo},
  \citenamefont {Go},\ and\ \citenamefont {Lee}}]{PRBJo2018}%
  \BibitemOpen
  \bibfield  {author} {\bibinfo {author} {\bibfnamefont {Daegeun}\ \bibnamefont
  {Jo}}, \bibinfo {author} {\bibfnamefont {Dongwook}\ \bibnamefont {Go}}, \
  and\ \bibinfo {author} {\bibfnamefont {Hyun-Woo}\ \bibnamefont {Lee}},\
  }\bibfield  {title} {\enquote {\bibinfo {title} {Gigantic intrinsic orbital
  hall effects in weakly spin-orbit coupled metals},}\ }\href {\doibase
  10.1103/PhysRevB.98.214405} {\bibfield  {journal} {\bibinfo  {journal} {Phys.
  Rev. B}\ }\textbf {\bibinfo {volume} {98}},\ \bibinfo {pages} {214405}
  (\bibinfo {year} {2018})}\BibitemShut {NoStop}%
\bibitem [{\citenamefont {Go}\ \emph {et~al.}(2018)\citenamefont {Go},
  \citenamefont {Jo}, \citenamefont {Kim},\ and\ \citenamefont
  {Lee}}]{PRLGo2018}%
  \BibitemOpen
  \bibfield  {author} {\bibinfo {author} {\bibfnamefont {Dongwook}\
  \bibnamefont {Go}}, \bibinfo {author} {\bibfnamefont {Daegeun}\ \bibnamefont
  {Jo}}, \bibinfo {author} {\bibfnamefont {Changyoung}\ \bibnamefont {Kim}}, \
  and\ \bibinfo {author} {\bibfnamefont {Hyun-Woo}\ \bibnamefont {Lee}},\
  }\bibfield  {title} {\enquote {\bibinfo {title} {Intrinsic spin and orbital
  hall effects from orbital texture},}\ }\href {\doibase
  10.1103/PhysRevLett.121.086602} {\bibfield  {journal} {\bibinfo  {journal}
  {Phys. Rev. Lett.}\ }\textbf {\bibinfo {volume} {121}},\ \bibinfo {pages}
  {086602} (\bibinfo {year} {2018})}\BibitemShut {NoStop}%
\bibitem [{\citenamefont {Park}\ \emph {et~al.}(2011)\citenamefont {Park},
  \citenamefont {Kim}, \citenamefont {Yu}, \citenamefont {Han},\ and\
  \citenamefont {Kim}}]{PRLPark2011}%
  \BibitemOpen
  \bibfield  {author} {\bibinfo {author} {\bibfnamefont {Seung~Ryong}\
  \bibnamefont {Park}}, \bibinfo {author} {\bibfnamefont {Choong~H.}\
  \bibnamefont {Kim}}, \bibinfo {author} {\bibfnamefont {Jaejun}\ \bibnamefont
  {Yu}}, \bibinfo {author} {\bibfnamefont {Jung~Hoon}\ \bibnamefont {Han}}, \
  and\ \bibinfo {author} {\bibfnamefont {Changyoung}\ \bibnamefont {Kim}},\
  }\bibfield  {title} {\enquote {\bibinfo {title} {Orbital-angular-momentum
  based origin of rashba-type surface band splitting},}\ }\href {\doibase
  10.1103/PhysRevLett.107.156803} {\bibfield  {journal} {\bibinfo  {journal}
  {Phys. Rev. Lett.}\ }\textbf {\bibinfo {volume} {107}},\ \bibinfo {pages}
  {156803} (\bibinfo {year} {2011})}\BibitemShut {NoStop}%
\bibitem [{\citenamefont {Park}\ \emph {et~al.}(2012)\citenamefont {Park},
  \citenamefont {Kim}, \citenamefont {Rhim},\ and\ \citenamefont
  {Han}}]{PRBPark2012}%
  \BibitemOpen
  \bibfield  {author} {\bibinfo {author} {\bibfnamefont {Jin-Hong}\
  \bibnamefont {Park}}, \bibinfo {author} {\bibfnamefont {Choong~H.}\
  \bibnamefont {Kim}}, \bibinfo {author} {\bibfnamefont {Jun-Won}\ \bibnamefont
  {Rhim}}, \ and\ \bibinfo {author} {\bibfnamefont {Jung~Hoon}\ \bibnamefont
  {Han}},\ }\bibfield  {title} {\enquote {\bibinfo {title} {Orbital rashba
  effect and its detection by circular dichroism angle-resolved photoemission
  spectroscopy},}\ }\href {\doibase 10.1103/PhysRevB.85.195401} {\bibfield
  {journal} {\bibinfo  {journal} {Phys. Rev. B}\ }\textbf {\bibinfo {volume}
  {85}},\ \bibinfo {pages} {195401} (\bibinfo {year} {2012})}\BibitemShut
  {NoStop}%
\bibitem [{\citenamefont {Bernevig}\ \emph {et~al.}(2005)\citenamefont
  {Bernevig}, \citenamefont {Hughes},\ and\ \citenamefont
  {Zhang}}]{PRLBernevig2005}%
  \BibitemOpen
  \bibfield  {author} {\bibinfo {author} {\bibfnamefont {B.~Andrei}\
  \bibnamefont {Bernevig}}, \bibinfo {author} {\bibfnamefont {Taylor~L.}\
  \bibnamefont {Hughes}}, \ and\ \bibinfo {author} {\bibfnamefont {Shou-Cheng}\
  \bibnamefont {Zhang}},\ }\bibfield  {title} {\enquote {\bibinfo {title}
  {Orbitronics: The intrinsic orbital current in $p$-doped silicon},}\ }\href
  {\doibase 10.1103/PhysRevLett.95.066601} {\bibfield  {journal} {\bibinfo
  {journal} {Phys. Rev. Lett.}\ }\textbf {\bibinfo {volume} {95}},\ \bibinfo
  {pages} {066601} (\bibinfo {year} {2005})}\BibitemShut {NoStop}%
\bibitem [{\citenamefont {Go}\ \emph {et~al.}(2017)\citenamefont {Go},
  \citenamefont {Hanke}, \citenamefont {Buhl}, \citenamefont {Freimuth},
  \citenamefont {Bihlmayer}, \citenamefont {Lee}, \citenamefont {Mokrousov},\
  and\ \citenamefont {Bl\"ugel}}]{SCIREPGo2017}%
  \BibitemOpen
  \bibfield  {author} {\bibinfo {author} {\bibfnamefont {Dongwook}\
  \bibnamefont {Go}}, \bibinfo {author} {\bibfnamefont {Jan-Philipp}\
  \bibnamefont {Hanke}}, \bibinfo {author} {\bibfnamefont {Patrick~M.}\
  \bibnamefont {Buhl}}, \bibinfo {author} {\bibfnamefont {Frank}\ \bibnamefont
  {Freimuth}}, \bibinfo {author} {\bibfnamefont {Gustav}\ \bibnamefont
  {Bihlmayer}}, \bibinfo {author} {\bibfnamefont {Hyun-Woo}\ \bibnamefont
  {Lee}}, \bibinfo {author} {\bibfnamefont {Yuriy}\ \bibnamefont {Mokrousov}},
  \ and\ \bibinfo {author} {\bibfnamefont {Stefan}\ \bibnamefont {Bl\"ugel}},\
  }\bibfield  {title} {\enquote {\bibinfo {title} {Toward surface orbitronics:
  giant orbital magnetism from the orbital rashba effect at the surface of
  sp-metals},}\ }\href {\doibase 10.1038/srep46742} {\bibfield  {journal}
  {\bibinfo  {journal} {Scientific Reports}\ }\textbf {\bibinfo {volume} {7}},\
  \bibinfo {pages} {46742} (\bibinfo {year} {2017})}\BibitemShut {NoStop}%
\bibitem [{\citenamefont {Hsu}\ \emph {et~al.}(2018)\citenamefont {Hsu},
  \citenamefont {Yao}, \citenamefont {Tan}, \citenamefont {Chang},
  \citenamefont {Liang},\ and\ \citenamefont {Jalil}}]{JPCMMing2018}%
  \BibitemOpen
  \bibfield  {author} {\bibinfo {author} {\bibfnamefont {Ming-Chien}\
  \bibnamefont {Hsu}}, \bibinfo {author} {\bibfnamefont {Liang-Zi}\
  \bibnamefont {Yao}}, \bibinfo {author} {\bibfnamefont {Seng~Ghee}\
  \bibnamefont {Tan}}, \bibinfo {author} {\bibfnamefont {Ching-Ray}\
  \bibnamefont {Chang}}, \bibinfo {author} {\bibfnamefont {Gengchiau}\
  \bibnamefont {Liang}}, \ and\ \bibinfo {author} {\bibfnamefont {Mansoor B~A}\
  \bibnamefont {Jalil}},\ }\bibfield  {title} {\enquote {\bibinfo {title}
  {Inherent orbital spin textures in rashba effect and their implications in
  spin-orbitronics},}\ }\href {\doibase 10.1088/1361-648x/aac86f} {\bibfield
  {journal} {\bibinfo  {journal} {Journal of Physics: Condensed Matter}\
  }\textbf {\bibinfo {volume} {30}},\ \bibinfo {pages} {285502} (\bibinfo
  {year} {2018})}\BibitemShut {NoStop}%
\bibitem [{\citenamefont {Manipatruni}\ \emph {et~al.}(2019)\citenamefont
  {Manipatruni}, \citenamefont {Nikonov}, \citenamefont {Lin}, \citenamefont
  {Gosavi}, \citenamefont {Liu}, \citenamefont {Prasad}, \citenamefont {Huang},
  \citenamefont {Bonturim}, \citenamefont {Ramesh},\ and\ \citenamefont
  {Young}}]{NATUREManipatruni2019}%
  \BibitemOpen
  \bibfield  {author} {\bibinfo {author} {\bibfnamefont {Sasikanth}\
  \bibnamefont {Manipatruni}}, \bibinfo {author} {\bibfnamefont {Dmitri~E.}\
  \bibnamefont {Nikonov}}, \bibinfo {author} {\bibfnamefont {Chia-Ching}\
  \bibnamefont {Lin}}, \bibinfo {author} {\bibfnamefont {Tanay~A.}\
  \bibnamefont {Gosavi}}, \bibinfo {author} {\bibfnamefont {Huichu}\
  \bibnamefont {Liu}}, \bibinfo {author} {\bibfnamefont {Bhagwati}\
  \bibnamefont {Prasad}}, \bibinfo {author} {\bibfnamefont {Yen-Lin}\
  \bibnamefont {Huang}}, \bibinfo {author} {\bibfnamefont {Everton}\
  \bibnamefont {Bonturim}}, \bibinfo {author} {\bibfnamefont {Ramamoorthy}\
  \bibnamefont {Ramesh}}, \ and\ \bibinfo {author} {\bibfnamefont {Ian~A.}\
  \bibnamefont {Young}},\ }\bibfield  {title} {\enquote {\bibinfo {title}
  {Scalable energy-efficient magnetoelectric spin-orbit logic},}\ }\href
  {\doibase 10.1038/s41586-018-0770-2} {\bibfield  {journal} {\bibinfo
  {journal} {Nature}\ }\textbf {\bibinfo {volume} {565}},\ \bibinfo {pages}
  {35--42} (\bibinfo {year} {2019})}\BibitemShut {NoStop}%
\bibitem [{\citenamefont {Chen}\ \emph {et~al.}(2018)\citenamefont {Chen},
  \citenamefont {Liu}, \citenamefont {Yang}, \citenamefont {Shi}, \citenamefont
  {Hu}, \citenamefont {Li},\ and\ \citenamefont {Zeng}}]{NATUREChen2018}%
  \BibitemOpen
  \bibfield  {author} {\bibinfo {author} {\bibfnamefont {Xi}~\bibnamefont
  {Chen}}, \bibinfo {author} {\bibfnamefont {Yang}\ \bibnamefont {Liu}},
  \bibinfo {author} {\bibfnamefont {Guang}\ \bibnamefont {Yang}}, \bibinfo
  {author} {\bibfnamefont {Hui}\ \bibnamefont {Shi}}, \bibinfo {author}
  {\bibfnamefont {Chen}\ \bibnamefont {Hu}}, \bibinfo {author} {\bibfnamefont
  {Minghua}\ \bibnamefont {Li}}, \ and\ \bibinfo {author} {\bibfnamefont
  {Haibo}\ \bibnamefont {Zeng}},\ }\bibfield  {title} {\enquote {\bibinfo
  {title} {Giant antidamping orbital torque originating from the orbital
  rashba-edelstein effect in ferromagnetic heterostructures},}\ }\href
  {\doibase 10.1038/s41467-018-05057-z} {\bibfield  {journal} {\bibinfo
  {journal} {Nature Communications}\ }\textbf {\bibinfo {volume} {9}},\
  \bibinfo {pages} {2569} (\bibinfo {year} {2018})}\BibitemShut {NoStop}%
\bibitem [{\citenamefont {Mannix}\ \emph {et~al.}(2015)\citenamefont {Mannix},
  \citenamefont {Zhou}, \citenamefont {Kiraly}, \citenamefont {Wood},
  \citenamefont {Alducin}, \citenamefont {Myers}, \citenamefont {Liu},
  \citenamefont {Fisher}, \citenamefont {Santiago}, \citenamefont {Guest},
  \citenamefont {Yacaman}, \citenamefont {Ponce}, \citenamefont {Oganov},
  \citenamefont {Hersam},\ and\ \citenamefont {Guisinger}}]{ScienceMannix2015}%
  \BibitemOpen
  \bibfield  {author} {\bibinfo {author} {\bibfnamefont {Andrew~J.}\
  \bibnamefont {Mannix}}, \bibinfo {author} {\bibfnamefont {Xiang-Feng}\
  \bibnamefont {Zhou}}, \bibinfo {author} {\bibfnamefont {Brian}\ \bibnamefont
  {Kiraly}}, \bibinfo {author} {\bibfnamefont {Joshua~D.}\ \bibnamefont
  {Wood}}, \bibinfo {author} {\bibfnamefont {Diego}\ \bibnamefont {Alducin}},
  \bibinfo {author} {\bibfnamefont {Benjamin~D.}\ \bibnamefont {Myers}},
  \bibinfo {author} {\bibfnamefont {Xiaolong}\ \bibnamefont {Liu}}, \bibinfo
  {author} {\bibfnamefont {Brandon~L.}\ \bibnamefont {Fisher}}, \bibinfo
  {author} {\bibfnamefont {Ulises}\ \bibnamefont {Santiago}}, \bibinfo {author}
  {\bibfnamefont {Jeffrey~R.}\ \bibnamefont {Guest}}, \bibinfo {author}
  {\bibfnamefont {Miguel~Jose}\ \bibnamefont {Yacaman}}, \bibinfo {author}
  {\bibfnamefont {Arturo}\ \bibnamefont {Ponce}}, \bibinfo {author}
  {\bibfnamefont {Artem~R.}\ \bibnamefont {Oganov}}, \bibinfo {author}
  {\bibfnamefont {Mark~C.}\ \bibnamefont {Hersam}}, \ and\ \bibinfo {author}
  {\bibfnamefont {Nathan~P.}\ \bibnamefont {Guisinger}},\ }\bibfield  {title}
  {\enquote {\bibinfo {title} {Synthesis of borophenes: Anisotropic,
  two-dimensional boron polymorphs},}\ }\href {\doibase
  10.1126/science.aad1080} {\bibfield  {journal} {\bibinfo  {journal}
  {Science}\ }\textbf {\bibinfo {volume} {350}},\ \bibinfo {pages} {1513--1516}
  (\bibinfo {year} {2015})}\BibitemShut {NoStop}%
\bibitem [{\citenamefont {Feng}\ \emph {et~al.}(2016)\citenamefont {Feng},
  \citenamefont {Zhang}, \citenamefont {Zhong}, \citenamefont {Li},
  \citenamefont {Li}, \citenamefont {Li}, \citenamefont {Cheng}, \citenamefont
  {Meng}, \citenamefont {Chen},\ and\ \citenamefont {Wu}}]{NatureChemFeng2016}%
  \BibitemOpen
  \bibfield  {author} {\bibinfo {author} {\bibfnamefont {Baojie}\ \bibnamefont
  {Feng}}, \bibinfo {author} {\bibfnamefont {Jin}\ \bibnamefont {Zhang}},
  \bibinfo {author} {\bibfnamefont {Qing}\ \bibnamefont {Zhong}}, \bibinfo
  {author} {\bibfnamefont {Wenbin}\ \bibnamefont {Li}}, \bibinfo {author}
  {\bibfnamefont {Shuai}\ \bibnamefont {Li}}, \bibinfo {author} {\bibfnamefont
  {Hui}\ \bibnamefont {Li}}, \bibinfo {author} {\bibfnamefont {Peng}\
  \bibnamefont {Cheng}}, \bibinfo {author} {\bibfnamefont {Sheng}\ \bibnamefont
  {Meng}}, \bibinfo {author} {\bibfnamefont {Lan}\ \bibnamefont {Chen}}, \ and\
  \bibinfo {author} {\bibfnamefont {Kehui}\ \bibnamefont {Wu}},\ }\bibfield
  {title} {\enquote {\bibinfo {title} {Experimental realization of
  two-dimensional boron sheets},}\ }\href {https://doi.org/10.1038/nchem.2491}
  {\bibfield  {journal} {\bibinfo  {journal} {Nature Chemistry}\ }\textbf
  {\bibinfo {volume} {8}},\ \bibinfo {pages} {563} (\bibinfo {year}
  {2016})}\BibitemShut {NoStop}%
\bibitem [{\citenamefont {Feng}\ \emph {et~al.}(2017)\citenamefont {Feng},
  \citenamefont {Sugino}, \citenamefont {Liu}, \citenamefont {Zhang},
  \citenamefont {Yukawa}, \citenamefont {Kawamura}, \citenamefont {Iimori},
  \citenamefont {Kim}, \citenamefont {Hasegawa}, \citenamefont {Li},
  \citenamefont {Chen}, \citenamefont {Wu}, \citenamefont {Kumigashira},
  \citenamefont {Komori}, \citenamefont {Chiang}, \citenamefont {Meng},\ and\
  \citenamefont {Matsuda}}]{PRLFeng2017}%
  \BibitemOpen
  \bibfield  {author} {\bibinfo {author} {\bibfnamefont {Baojie}\ \bibnamefont
  {Feng}}, \bibinfo {author} {\bibfnamefont {Osamu}\ \bibnamefont {Sugino}},
  \bibinfo {author} {\bibfnamefont {Ro-Ya}\ \bibnamefont {Liu}}, \bibinfo
  {author} {\bibfnamefont {Jin}\ \bibnamefont {Zhang}}, \bibinfo {author}
  {\bibfnamefont {Ryu}\ \bibnamefont {Yukawa}}, \bibinfo {author}
  {\bibfnamefont {Mitsuaki}\ \bibnamefont {Kawamura}}, \bibinfo {author}
  {\bibfnamefont {Takushi}\ \bibnamefont {Iimori}}, \bibinfo {author}
  {\bibfnamefont {Howon}\ \bibnamefont {Kim}}, \bibinfo {author} {\bibfnamefont
  {Yukio}\ \bibnamefont {Hasegawa}}, \bibinfo {author} {\bibfnamefont {Hui}\
  \bibnamefont {Li}}, \bibinfo {author} {\bibfnamefont {Lan}\ \bibnamefont
  {Chen}}, \bibinfo {author} {\bibfnamefont {Kehui}\ \bibnamefont {Wu}},
  \bibinfo {author} {\bibfnamefont {Hiroshi}\ \bibnamefont {Kumigashira}},
  \bibinfo {author} {\bibfnamefont {Fumio}\ \bibnamefont {Komori}}, \bibinfo
  {author} {\bibfnamefont {Tai-Chang}\ \bibnamefont {Chiang}}, \bibinfo
  {author} {\bibfnamefont {Sheng}\ \bibnamefont {Meng}}, \ and\ \bibinfo
  {author} {\bibfnamefont {Iwao}\ \bibnamefont {Matsuda}},\ }\bibfield  {title}
  {\enquote {\bibinfo {title} {Dirac fermions in borophene},}\ }\href {\doibase
  10.1103/PhysRevLett.118.096401} {\bibfield  {journal} {\bibinfo  {journal}
  {Phys. Rev. Lett.}\ }\textbf {\bibinfo {volume} {118}},\ \bibinfo {pages}
  {096401} (\bibinfo {year} {2017})}\BibitemShut {NoStop}%
\bibitem [{\citenamefont {Xu}\ \emph {et~al.}(2016)\citenamefont {Xu},
  \citenamefont {Zhao}, \citenamefont {Liao}, \citenamefont {Yang},\ and\
  \citenamefont {Xu}}]{NanoRXu2016}%
  \BibitemOpen
  \bibfield  {author} {\bibinfo {author} {\bibfnamefont {Shaogang}\
  \bibnamefont {Xu}}, \bibinfo {author} {\bibfnamefont {Yujun}\ \bibnamefont
  {Zhao}}, \bibinfo {author} {\bibfnamefont {Jihai}\ \bibnamefont {Liao}},
  \bibinfo {author} {\bibfnamefont {Xiaobao}\ \bibnamefont {Yang}}, \ and\
  \bibinfo {author} {\bibfnamefont {Hu}~\bibnamefont {Xu}},\ }\bibfield
  {title} {\enquote {\bibinfo {title} {The nucleation and growth of borophene
  on the ag (111) surface},}\ }\href {\doibase 10.1007/s12274-016-1148-0}
  {\bibfield  {journal} {\bibinfo  {journal} {Nano Research}\ }\textbf
  {\bibinfo {volume} {9}},\ \bibinfo {pages} {2616--2622} (\bibinfo {year}
  {2016})}\BibitemShut {NoStop}%
\bibitem [{\citenamefont {Yi}\ \emph {et~al.}(2017)\citenamefont {Yi},
  \citenamefont {Liu}, \citenamefont {Botana}, \citenamefont {Zhao},
  \citenamefont {Liu}, \citenamefont {Liu},\ and\ \citenamefont
  {Miao}}]{JPCLYi2017}%
  \BibitemOpen
  \bibfield  {author} {\bibinfo {author} {\bibfnamefont {Wen-cai}\ \bibnamefont
  {Yi}}, \bibinfo {author} {\bibfnamefont {Wei}\ \bibnamefont {Liu}}, \bibinfo
  {author} {\bibfnamefont {Jorge}\ \bibnamefont {Botana}}, \bibinfo {author}
  {\bibfnamefont {Lei}\ \bibnamefont {Zhao}}, \bibinfo {author} {\bibfnamefont
  {Zhen}\ \bibnamefont {Liu}}, \bibinfo {author} {\bibfnamefont {Jing-yao}\
  \bibnamefont {Liu}}, \ and\ \bibinfo {author} {\bibfnamefont {Mao-sheng}\
  \bibnamefont {Miao}},\ }\bibfield  {title} {\enquote {\bibinfo {title}
  {Honeycomb boron allotropes with dirac cones: A true analogue to graphene},}\
  }\href {\doibase 10.1021/acs.jpclett.7b00891} {\bibfield  {journal} {\bibinfo
   {journal} {The Journal of Physical Chemistry Letters}\ }\textbf {\bibinfo
  {volume} {8}},\ \bibinfo {pages} {2647--2653} (\bibinfo {year}
  {2017})}\BibitemShut {NoStop}%
\bibitem [{\citenamefont {Kong}\ \emph {et~al.}(2018)\citenamefont {Kong},
  \citenamefont {Wu},\ and\ \citenamefont {Chen}}]{FPFong2018}%
  \BibitemOpen
  \bibfield  {author} {\bibinfo {author} {\bibfnamefont {Longjuan}\
  \bibnamefont {Kong}}, \bibinfo {author} {\bibfnamefont {Kehui}\ \bibnamefont
  {Wu}}, \ and\ \bibinfo {author} {\bibfnamefont {Lan}\ \bibnamefont {Chen}},\
  }\bibfield  {title} {\enquote {\bibinfo {title} {Recent progress on
  borophene: Growth and structures},}\ }\href {\doibase
  10.1007/s11467-018-0752-8} {\bibfield  {journal} {\bibinfo  {journal}
  {Frontiers of Physics}\ }\textbf {\bibinfo {volume} {13}},\ \bibinfo {pages}
  {138105} (\bibinfo {year} {2018})}\BibitemShut {NoStop}%
\bibitem [{\citenamefont {Grunbaum}\ and\ \citenamefont
  {Shephard}(1977)}]{MATBranko1977}%
  \BibitemOpen
  \bibfield  {author} {\bibinfo {author} {\bibfnamefont {Branko}\ \bibnamefont
  {Grunbaum}}\ and\ \bibinfo {author} {\bibfnamefont {Geoffrey~C.}\
  \bibnamefont {Shephard}},\ }\bibfield  {title} {\enquote {\bibinfo {title}
  {Tilings by regular polygons},}\ }\href {http://www.jstor.org/stable/2689529}
  {\bibfield  {journal} {\bibinfo  {journal} {Mathematics Magazine}\ }\textbf
  {\bibinfo {volume} {50}},\ \bibinfo {pages} {227--247} (\bibinfo {year}
  {1977})}\BibitemShut {NoStop}%
\bibitem [{\citenamefont {Chang}\ \emph {et~al.}(2017)\citenamefont {Chang},
  \citenamefont {Xu}, \citenamefont {Wieder}, \citenamefont {Sanchez},
  \citenamefont {Huang}, \citenamefont {Belopolski}, \citenamefont {Chang},
  \citenamefont {Zhang}, \citenamefont {Bansil}, \citenamefont {Lin},\ and\
  \citenamefont {Hasan}}]{PRLChang2017}%
  \BibitemOpen
  \bibfield  {author} {\bibinfo {author} {\bibfnamefont {Guoqing}\ \bibnamefont
  {Chang}}, \bibinfo {author} {\bibfnamefont {Su-Yang}\ \bibnamefont {Xu}},
  \bibinfo {author} {\bibfnamefont {Benjamin~J.}\ \bibnamefont {Wieder}},
  \bibinfo {author} {\bibfnamefont {Daniel~S.}\ \bibnamefont {Sanchez}},
  \bibinfo {author} {\bibfnamefont {Shin-Ming}\ \bibnamefont {Huang}}, \bibinfo
  {author} {\bibfnamefont {Ilya}\ \bibnamefont {Belopolski}}, \bibinfo {author}
  {\bibfnamefont {Tay-Rong}\ \bibnamefont {Chang}}, \bibinfo {author}
  {\bibfnamefont {Songtian}\ \bibnamefont {Zhang}}, \bibinfo {author}
  {\bibfnamefont {Arun}\ \bibnamefont {Bansil}}, \bibinfo {author}
  {\bibfnamefont {Hsin}\ \bibnamefont {Lin}}, \ and\ \bibinfo {author}
  {\bibfnamefont {M.~Zahid}\ \bibnamefont {Hasan}},\ }\bibfield  {title}
  {\enquote {\bibinfo {title} {Unconventional chiral fermions and large
  topological fermi arcs in rhsi},}\ }\href {\doibase
  10.1103/PhysRevLett.119.206401} {\bibfield  {journal} {\bibinfo  {journal}
  {Phys. Rev. Lett.}\ }\textbf {\bibinfo {volume} {119}},\ \bibinfo {pages}
  {206401} (\bibinfo {year} {2017})}\BibitemShut {NoStop}%
\bibitem [{\citenamefont {Chang}\ \emph {et~al.}(2018)\citenamefont {Chang},
  \citenamefont {Wieder}, \citenamefont {Schindler}, \citenamefont {Sanchez},
  \citenamefont {Belopolski}, \citenamefont {Huang}, \citenamefont {Singh},
  \citenamefont {Wu}, \citenamefont {Chang}, \citenamefont {Neupert},
  \citenamefont {Xu}, \citenamefont {Lin},\ and\ \citenamefont
  {Hasan}}]{NATUREChang2018}%
  \BibitemOpen
  \bibfield  {author} {\bibinfo {author} {\bibfnamefont {Guoqing}\ \bibnamefont
  {Chang}}, \bibinfo {author} {\bibfnamefont {Benjamin~J.}\ \bibnamefont
  {Wieder}}, \bibinfo {author} {\bibfnamefont {Frank}\ \bibnamefont
  {Schindler}}, \bibinfo {author} {\bibfnamefont {Daniel~S.}\ \bibnamefont
  {Sanchez}}, \bibinfo {author} {\bibfnamefont {Ilya}\ \bibnamefont
  {Belopolski}}, \bibinfo {author} {\bibfnamefont {Shin-Ming}\ \bibnamefont
  {Huang}}, \bibinfo {author} {\bibfnamefont {Bahadur}\ \bibnamefont {Singh}},
  \bibinfo {author} {\bibfnamefont {Di}~\bibnamefont {Wu}}, \bibinfo {author}
  {\bibfnamefont {Tay-Rong}\ \bibnamefont {Chang}}, \bibinfo {author}
  {\bibfnamefont {Titus}\ \bibnamefont {Neupert}}, \bibinfo {author}
  {\bibfnamefont {Su-Yang}\ \bibnamefont {Xu}}, \bibinfo {author}
  {\bibfnamefont {Hsin}\ \bibnamefont {Lin}}, \ and\ \bibinfo {author}
  {\bibfnamefont {M.~Zahid}\ \bibnamefont {Hasan}},\ }\bibfield  {title}
  {\enquote {\bibinfo {title} {Topological quantum properties of chiral
  crystals},}\ }\href {\doibase 10.1038/s41563-018-0169-3} {\bibfield
  {journal} {\bibinfo  {journal} {Nature Materials}\ }\textbf {\bibinfo
  {volume} {17}},\ \bibinfo {pages} {978--985} (\bibinfo {year}
  {2018})}\BibitemShut {NoStop}%
\bibitem [{\citenamefont {Hanakata}\ \emph {et~al.}(2017)\citenamefont
  {Hanakata}, \citenamefont {Rodin}, \citenamefont {Carvalho}, \citenamefont
  {Park}, \citenamefont {Campbell},\ and\ \citenamefont
  {Castro~Neto}}]{PRBHanakata2017}%
  \BibitemOpen
  \bibfield  {author} {\bibinfo {author} {\bibfnamefont {Paul~Z.}\ \bibnamefont
  {Hanakata}}, \bibinfo {author} {\bibfnamefont {A.~S.}\ \bibnamefont {Rodin}},
  \bibinfo {author} {\bibfnamefont {Alexandra}\ \bibnamefont {Carvalho}},
  \bibinfo {author} {\bibfnamefont {Harold~S.}\ \bibnamefont {Park}}, \bibinfo
  {author} {\bibfnamefont {David~K.}\ \bibnamefont {Campbell}}, \ and\ \bibinfo
  {author} {\bibfnamefont {A.~H.}\ \bibnamefont {Castro~Neto}},\ }\bibfield
  {title} {\enquote {\bibinfo {title} {Two-dimensional square buckled rashba
  lead chalcogenides},}\ }\href {\doibase 10.1103/PhysRevB.96.161401}
  {\bibfield  {journal} {\bibinfo  {journal} {Phys. Rev. B}\ }\textbf {\bibinfo
  {volume} {96}},\ \bibinfo {pages} {161401} (\bibinfo {year}
  {2017})}\BibitemShut {NoStop}%
\bibitem [{\citenamefont {Alexandradinata}\ and\ \citenamefont
  {Bernevig}(2015)}]{PSAlexandradinata2015}%
  \BibitemOpen
  \bibfield  {author} {\bibinfo {author} {\bibfnamefont {A}~\bibnamefont
  {Alexandradinata}}\ and\ \bibinfo {author} {\bibfnamefont {B~Andrei}\
  \bibnamefont {Bernevig}},\ }\bibfield  {title} {\enquote {\bibinfo {title}
  {Spin-orbit-free topological insulators},}\ }\href {\doibase
  10.1088/0031-8949/2015/t164/014013} {\bibfield  {journal} {\bibinfo
  {journal} {Physica Scripta}\ }\textbf {\bibinfo {volume} {T164}},\ \bibinfo
  {pages} {014013} (\bibinfo {year} {2015})}\BibitemShut {NoStop}%
\bibitem [{\citenamefont {Ferreira}\ and\ \citenamefont
  {Loss}(2013)}]{PRLFerreira2013}%
  \BibitemOpen
  \bibfield  {author} {\bibinfo {author} {\bibfnamefont {Gerson~J.}\
  \bibnamefont {Ferreira}}\ and\ \bibinfo {author} {\bibfnamefont {Daniel}\
  \bibnamefont {Loss}},\ }\bibfield  {title} {\enquote {\bibinfo {title}
  {Magnetically defined qubits on 3d topological insulators},}\ }\href
  {\doibase 10.1103/PhysRevLett.111.106802} {\bibfield  {journal} {\bibinfo
  {journal} {Phys. Rev. Lett.}\ }\textbf {\bibinfo {volume} {111}},\ \bibinfo
  {pages} {106802} (\bibinfo {year} {2013})}\BibitemShut {NoStop}%
\bibitem [{\citenamefont {Shen}\ \emph {et~al.}(2014)\citenamefont {Shen},
  \citenamefont {Vignale},\ and\ \citenamefont {Raimondi}}]{PRLShen2014}%
  \BibitemOpen
  \bibfield  {author} {\bibinfo {author} {\bibfnamefont {Ka}~\bibnamefont
  {Shen}}, \bibinfo {author} {\bibfnamefont {G.}~\bibnamefont {Vignale}}, \
  and\ \bibinfo {author} {\bibfnamefont {R.}~\bibnamefont {Raimondi}},\
  }\bibfield  {title} {\enquote {\bibinfo {title} {Microscopic theory of the
  inverse edelstein effect},}\ }\href {\doibase 10.1103/PhysRevLett.112.096601}
  {\bibfield  {journal} {\bibinfo  {journal} {Phys. Rev. Lett.}\ }\textbf
  {\bibinfo {volume} {112}},\ \bibinfo {pages} {096601} (\bibinfo {year}
  {2014})}\BibitemShut {NoStop}%
\bibitem [{\citenamefont {Rojas-S\'anchez}\ \emph {et~al.}(2016)\citenamefont
  {Rojas-S\'anchez}, \citenamefont {Oyarz\'un}, \citenamefont {Fu},
  \citenamefont {Marty}, \citenamefont {Vergnaud}, \citenamefont {Gambarelli},
  \citenamefont {Vila}, \citenamefont {Jamet}, \citenamefont {Ohtsubo},
  \citenamefont {Taleb-Ibrahimi}, \citenamefont {Le~F\`evre}, \citenamefont
  {Bertran}, \citenamefont {Reyren}, \citenamefont {George},\ and\
  \citenamefont {Fert}}]{PRLRojas2016}%
  \BibitemOpen
  \bibfield  {author} {\bibinfo {author} {\bibfnamefont {J.-C.}\ \bibnamefont
  {Rojas-S\'anchez}}, \bibinfo {author} {\bibfnamefont {S.}~\bibnamefont
  {Oyarz\'un}}, \bibinfo {author} {\bibfnamefont {Y.}~\bibnamefont {Fu}},
  \bibinfo {author} {\bibfnamefont {A.}~\bibnamefont {Marty}}, \bibinfo
  {author} {\bibfnamefont {C.}~\bibnamefont {Vergnaud}}, \bibinfo {author}
  {\bibfnamefont {S.}~\bibnamefont {Gambarelli}}, \bibinfo {author}
  {\bibfnamefont {L.}~\bibnamefont {Vila}}, \bibinfo {author} {\bibfnamefont
  {M.}~\bibnamefont {Jamet}}, \bibinfo {author} {\bibfnamefont
  {Y.}~\bibnamefont {Ohtsubo}}, \bibinfo {author} {\bibfnamefont
  {A.}~\bibnamefont {Taleb-Ibrahimi}}, \bibinfo {author} {\bibfnamefont
  {P.}~\bibnamefont {Le~F\`evre}}, \bibinfo {author} {\bibfnamefont
  {F.}~\bibnamefont {Bertran}}, \bibinfo {author} {\bibfnamefont
  {N.}~\bibnamefont {Reyren}}, \bibinfo {author} {\bibfnamefont {J.-M.}\
  \bibnamefont {George}}, \ and\ \bibinfo {author} {\bibfnamefont
  {A.}~\bibnamefont {Fert}},\ }\bibfield  {title} {\enquote {\bibinfo {title}
  {Spin to charge conversion at room temperature by spin pumping into a new
  type of topological insulator: $\ensuremath{\alpha}$-sn films},}\ }\href
  {\doibase 10.1103/PhysRevLett.116.096602} {\bibfield  {journal} {\bibinfo
  {journal} {Phys. Rev. Lett.}\ }\textbf {\bibinfo {volume} {116}},\ \bibinfo
  {pages} {096602} (\bibinfo {year} {2016})}\BibitemShut {NoStop}%
\bibitem [{\citenamefont {Crasto~de Lima}\ \emph {et~al.}(2019)\citenamefont
  {Crasto~de Lima}, \citenamefont {Ferreira},\ and\ \citenamefont
  {Miwa}}]{JCPCrasto2019}%
  \BibitemOpen
  \bibfield  {author} {\bibinfo {author} {\bibfnamefont {F.}~\bibnamefont
  {Crasto~de Lima}}, \bibinfo {author} {\bibfnamefont {G.~J.}\ \bibnamefont
  {Ferreira}}, \ and\ \bibinfo {author} {\bibfnamefont {R.~H.}\ \bibnamefont
  {Miwa}},\ }\bibfield  {title} {\enquote {\bibinfo {title} {Layertronic
  control of topological states in multilayer metal-organic frameworks},}\
  }\href {\doibase 10.1063/1.5100679} {\bibfield  {journal} {\bibinfo
  {journal} {The Journal of Chemical Physics}\ }\textbf {\bibinfo {volume}
  {150}},\ \bibinfo {pages} {234701} (\bibinfo {year} {2019})}\BibitemShut
  {NoStop}%
\bibitem [{\citenamefont {de~Lima}\ \emph {et~al.}(2017)\citenamefont
  {de~Lima}, \citenamefont {Ferreira},\ and\ \citenamefont
  {Miwa}}]{PRBdeLima2017}%
  \BibitemOpen
  \bibfield  {author} {\bibinfo {author} {\bibfnamefont {F.~Crasto}\
  \bibnamefont {de~Lima}}, \bibinfo {author} {\bibfnamefont {Gerson~J.}\
  \bibnamefont {Ferreira}}, \ and\ \bibinfo {author} {\bibfnamefont {R.~H.}\
  \bibnamefont {Miwa}},\ }\bibfield  {title} {\enquote {\bibinfo {title}
  {Tuning the topological states in metal-organic bilayers},}\ }\href {\doibase
  10.1103/PhysRevB.96.115426} {\bibfield  {journal} {\bibinfo  {journal} {Phys.
  Rev. B}\ }\textbf {\bibinfo {volume} {96}},\ \bibinfo {pages} {115426}
  (\bibinfo {year} {2017})}\BibitemShut {NoStop}%
\bibitem [{not()}]{note1}%
  \BibitemOpen
  \href@noop {} {}\bibinfo {note} {The pseudospinor $\sigma_x = \ket{+}\bra{-}
  + \ket{-}\bra{+} \equiv \ket{p_x}\bra{p_x} - \ket{p_y} \bra{p_y}$ mean value
  is easily obtained from DFT as $\avg{\sigma_x} = |\avg{p_x | \Psi}|^2 - |
  \avg{p_y | \Psi}|^2$, with $\ket{\Psi}$ been the eigenvector obtained through
  the DFT calculation.}\BibitemShut {Stop}%
\bibitem [{\citenamefont {Shan}\ \emph {et~al.}(2010)\citenamefont {Shan},
  \citenamefont {Lu},\ and\ \citenamefont {Shen}}]{NJPShan2010}%
  \BibitemOpen
  \bibfield  {author} {\bibinfo {author} {\bibfnamefont {Wen-Yu}\ \bibnamefont
  {Shan}}, \bibinfo {author} {\bibfnamefont {Hai-Zhou}\ \bibnamefont {Lu}}, \
  and\ \bibinfo {author} {\bibfnamefont {Shun-Qing}\ \bibnamefont {Shen}},\
  }\bibfield  {title} {\enquote {\bibinfo {title} {Effective continuous model
  for surface states and thin films of three-dimensional topological
  insulators},}\ }\href {\doibase 10.1088/1367-2630/12/4/043048} {\bibfield
  {journal} {\bibinfo  {journal} {New Journal of Physics}\ }\textbf {\bibinfo
  {volume} {12}},\ \bibinfo {pages} {043048} (\bibinfo {year}
  {2010})}\BibitemShut {NoStop}%
\bibitem [{\citenamefont {Kresse}\ and\ \citenamefont
  {Furthmüller}(1996)}]{VASP}%
  \BibitemOpen
  \bibfield  {author} {\bibinfo {author} {\bibfnamefont {G.}~\bibnamefont
  {Kresse}}\ and\ \bibinfo {author} {\bibfnamefont {J.}~\bibnamefont
  {Furthmüller}},\ }\bibfield  {title} {\enquote {\bibinfo {title} {Efficiency
  of ab-initio total energy calculations for metals and semiconductors using a
  plane-wave basis set},}\ }\href {\doibase
  https://doi.org/10.1016/0927-0256(96)00008-0} {\bibfield  {journal} {\bibinfo
   {journal} {Computational Materials Science}\ }\textbf {\bibinfo {volume}
  {6}},\ \bibinfo {pages} {15 -- 50} (\bibinfo {year} {1996})}\BibitemShut
  {NoStop}%
\bibitem [{\citenamefont {Perdew}\ \emph {et~al.}(1996)\citenamefont {Perdew},
  \citenamefont {Burke},\ and\ \citenamefont {Ernzerhof}}]{PBE}%
  \BibitemOpen
  \bibfield  {author} {\bibinfo {author} {\bibfnamefont {J.~P.}\ \bibnamefont
  {Perdew}}, \bibinfo {author} {\bibfnamefont {K.}~\bibnamefont {Burke}}, \
  and\ \bibinfo {author} {\bibfnamefont {M.}~\bibnamefont {Ernzerhof}},\
  }\bibfield  {title} {\enquote {\bibinfo {title} {Generalized gradient
  approximation made simple},}\ }\href@noop {} {\bibfield  {journal} {\bibinfo
  {journal} {Phys. Rev. Lett.}\ }\textbf {\bibinfo {volume} {77}},\ \bibinfo
  {pages} {3865} (\bibinfo {year} {1996})}\BibitemShut {NoStop}%
\bibitem [{\citenamefont {Monkhorst}\ and\ \citenamefont
  {Pack}(1976)}]{PhysRevB.13.5188}%
  \BibitemOpen
  \bibfield  {author} {\bibinfo {author} {\bibfnamefont {Hendrik~J.}\
  \bibnamefont {Monkhorst}}\ and\ \bibinfo {author} {\bibfnamefont {James~D.}\
  \bibnamefont {Pack}},\ }\bibfield  {title} {\enquote {\bibinfo {title}
  {Special points for brillouin-zone integrations},}\ }\href {\doibase
  10.1103/PhysRevB.13.5188} {\bibfield  {journal} {\bibinfo  {journal} {Phys.
  Rev. B}\ }\textbf {\bibinfo {volume} {13}},\ \bibinfo {pages} {5188--5192}
  (\bibinfo {year} {1976})}\BibitemShut {NoStop}%
\bibitem [{\citenamefont {Bl\"ochl}(1994)}]{PAW}%
  \BibitemOpen
  \bibfield  {author} {\bibinfo {author} {\bibfnamefont {P.~E.}\ \bibnamefont
  {Bl\"ochl}},\ }\bibfield  {title} {\enquote {\bibinfo {title} {Projector
  augmented-wave method},}\ }\href {\doibase 10.1103/PhysRevB.50.17953}
  {\bibfield  {journal} {\bibinfo  {journal} {Phys. Rev. B}\ }\textbf {\bibinfo
  {volume} {50}},\ \bibinfo {pages} {17953--17979} (\bibinfo {year}
  {1994})}\BibitemShut {NoStop}%
\end{thebibliography}%

\end{document}